# **Synergy or Rivalry?** Glimpses of Regional Modernization and Public Service Equalization: A Case Study from China

Shengwen Shi [a,1], Jian'an Zhang [b,*,1]

a. School of Economics, Shanghai University of Finance and Economics, Shanghai 200433, China
b. School of Mathematics, Shanghai University of Finance and Economics, Shanghai 200433, China

**Abstract:** For most developing countries, increasing the equalization of basic public services is widely recognized as an effective channel to improve people's sense of contentment. However, for many emerging economies like China, the equalization level of basic public services may often be neglected in the trade-off between the speed and quality of development. Taking the Yangtze River Delta region of China as an example, this paper first adopts the coupling coordination degree model to explore current status of basic public services in this region, and then uses Moran's I index to study the overall equalization level of development there. Moreover, this paper uses the Theil index to analyze the main reasons for the spatial differences in the level of public services, followed by the AF method to accurately identify the exact weaknesses of the 40 counties of 10 cities with the weakest level of basic public service development. Based on this, this paper provides targeted optimization initiatives and continues to explore the factors affecting the growth of the level of public service equalization through the convergence model, verifying the convergence trend of the degree of public service equalization, and ultimately providing practical policy recommendations for promoting the equalization of basic public services.

## 1. Introduction

As an important factor that effectively affects people's sense of contentment and happiness, equalization of basic public services is generally regarded as an important means to achieve the goal of common prosperity (Cheng & Liu, 2023). Over the past few decades, however, development goals in many countries have been limited to increasing incomes while neglecting the provision and optimization of public services (Anand & Ravallion, 1993). Fortunately, in recent years, many developing countries are continuously improving the supply of basic public services and enhancing the level of equality of basic public services during the process of economic transformation and upgrading from high-speed development to high-quality development, which will be an important breakthrough in promoting high-quality development and regional modernization.

To measure the performance of basic public services, there are usually three major criteria: efficiency, effectiveness and equality, among which the importance of equity is becoming increasingly prominent (Savas, 1978). Since Dwight Waldo introduced the new government perspective, equality has played a key role in public administration and public policy research. However, many studies focus on employment, politics, jurisprudence, voting and other issues, but ignore the role of public services (Cepiku & Mastrodascio, 2021). In fact, considering equality in the issue of public service delivery can bring considerable benefits (Akoluk & Karsu, 2022). Practice has proved that citizens can and do exert significant influence on public policy by participating in the implementation of public projects (Whitaker, 1980). In many developed countries such as the United States, local governments have adopted initiatives like 311 interactive platforms to promote distributive equity in public service delivery by encouraging citizens to participate (Xu & Tang, 2020). However, in many developing countries, the provision of development-oriented public services still has not become consensus. In South Africa, for example, policies and strategies aimed at dealing with inequality, poverty, and education have not been effectively implemented (Shabangu & Madzivhandila, 2017).

Policy and governance have become central to explaining the widespread inadequacy of public services in developing countries (Batley, McCourt & Mcloughlin, 2012). Therefore, governments and the public in developing countries must confront regional disparities that pose challenges to the economy and public administration. The empirical results based on the spatial econometric model show that the equalization of various public services can promote the equalization of regional income and consumption (Li et al., 2017). This paper takes China, the world's largest developing country, as an example. Inadequate equalization of public services and the coexistence of high investment and low consumption have always been the two major dilemmas hindering China's high-quality economic development (Dai et al., 2022) The rapid development of China's urban economy has attracted more and more people to flood in cities. However, in the rapidly urbanizing Chinese landscape, efforts still need to be made to further demonstrate social equity and inclusiveness in public services. (Ouyang et al., 2017)

However, as the definition of equalization itself reveals, the realization of equalization in basic public services is bound to involve cooperation both within and between regions. In the era of globalization, the connections between cities are more and more complex and diverse, and the coordinated development of regional economy is increasingly important, which will also bring a significant impact on the realization of the equalization of basic public services. (Ye et al., 2019) In recent years, inter-city cooperation in China's Yangtze River Delta (YRD) region has attracted much attention from both the government and academia. (Luo & Shen, 2009) Meanwhile, the analysis of regional cooperation in the Yangtze River Delta region is widely regarded as an important entry point to understand China's urban regional development. (Li & Wu, 2018) Therefore, selecting the Yangtze River Delta region to analyze the correlation between the equalization of basic public services and modernization has certain academic and practical value.

In previous researches, scholars have used system generalized method of moments (SYS-GMM), intermediary effect, linkage effect and many other methods to conduct empirical research from multiple dimensions such as government governance, medical and health care, public education, environmental protection, public culture and social organization density (Yang, Xue & Ma, 2019). Solutions have also been provided to improve the level of basic public services from the perspectives of smart city construction (Zhou, Liu & Wang, 2022) and the management mode of the Internet of Things (Wu & Xiao, 2022). However, in general, detailed quantitative studies on modernization and the equalization of basic public services are relatively lacking. To this end, this paper will take China's Yangtze River Delta region as an example, providing quantitative suggestions for developing countries to better promote the equalization of basic public services and modernization by comprehensively applying the coupling coordination degree model, Moran index, Theil index, AF model, $\beta$ convergence model and other quantitative methods.

**Compare with the existing studies, this paper shows following aspects of innovation through the entire research process.**

Initially, this paper is the first of its kind to construct a quantitative measurement system that combines time and space, with the aim of accurately measuring the level of equalization of basic public services. Based on an in-depth exploration of the existing literature, we ensure the comprehensiveness of the measurement and construct an assessment framework of seven first-level indicators to highlight its completeness and breadth.

In data processing, we paid especial attention to the impact of extreme values and adopted the efficacy function in the form of power function model to achieve the nondimensionalization of data. This method not only improves the observation of the differences in regional development levels, but also clearly highlights the differences in nature, making the development gap more significant and easier to analyze.

Meanwhile, in the time dimension, we use the coupling coordination degree model for in-depth analysis, while in the spatial dimension, we further extend the existing basic public service index measures. Particularly, we adopt the Theil index for inter-regional variability analysis, based on which we not merely deepen our exploration of the variability in the scores of the first-level indicators of basic public service equalization across regions in 2021, but also provide a critical insight into the root causes of these differences, delivering valuable suggestions into the direction of advancing public service equalization.

Moreover, we also introduce the AF model and the $\beta$ convergence model. The AF method is widely used in multidimensional poverty measurement, which allows for the convergence of multidimensional indicators into a single comparable index and has strong policy analysis advantages. Considering the similarity between poverty issues and the basic public service equalization problem in this study, we decided to adopt the AF model to analyze the shortcomings of public services in each region in depth. In addition, based on the theory of regional economic convergence, we introduced the convergence model to explore the dynamic development trend of basic public services. We try to determine whether the level of public services is moving towards a steady state of equalization, and further, explore whether there is a possibility for regions with low level of basic public services to catch up with high level regions during the process of modernization.

## 2. Data Collection and Processing

### 2.1 Data collection path

In order to measure the level of basic public services in the Yangtze River Delta region more objectively and accurately, this paper comprehensively selects data from several authoritative databases, including the China Urban Statistical Yearbook, the CECN statistical database and the CSMAR database. By integrating five years of raw datasets, we paid particular attention to ensuring the timeliness and continuity of the data, thus enabling our follow-up comparative analysis of trends and changes between years whilst providing deeper insights into the progress of public service equalization. Such a time span enhances the depth and breadth of our study, making the conclusions more precise and reliable.

### 2.2 Data processing

#### 2.2.1 Interpolation and averaging
In the process of data preprocessing, this paper adopts the interpolation method and the mean value method to deal with individual missing values to ensure the completeness and accuracy of the data. Through longitudinal data collection, this paper realizes analyzing the situation of each city in different years, so as to observe the trend of changes in the level of basic public services more comprehensively.

#### 2.2.2 Inverse Indicator Reversal
When dealing with indicator data, this paper pays special attention to the existence of inverse indicators. Reverse indicators need to be inverted when calculating indicator scores to ensure that the larger the indicator value indicates a better situation. This paper ensures that all indicators are positive indicators when calculating scores by processing reverse indicators to guarantee the consistency and comparability of indicator scores.

#### 2.2.3 Nondimensionalization
In order to compare indicators of different natures and units of measurement, this paper combines the numerical characteristics of the indicators of basic public service level and chooses the power function type efficacy function to carry out the dimensionless treatment. This treatment can normalize the data of different indicators, making them comparable, thus enabling better observation of the development differences among regions and highlighting the contribution of different indicators in the overall indicator scores.

The power function type efficacy function is formulated as follows:

$$d = \frac{x^2 - x_l^2}{x_h^2 - x_l^2} \times 100 \qquad (1)$$

#### 2.2.4 Abnormal extremes processing
Meanwhile, to mitigate the impact of abnormal extremes that may occur under each indicator, this paper takes the 95% quantile of the actual value of the indicator data as the upper limit and the 5% quantile as the lower limit.

After getting the percentage scores of each indicator, the entropy weight method of objective assignment is adopted to assign the indicators. And finally, the quantitative score of the basic public service level of each city in five years is calculated.

## 3. Building the evaluation system for basic public service indicators

In order to explore the current situation of basic public services and its inter-regional differences in depth, authors of this paper conducted field survey of the Yangtze River Delta region to more precisely grasp the state of public services in each province or city. Based on the differences captured in the field observations, this paper then develops a series of specific research indicators. These proposed indicators cover a number of key areas, such as resource allocation, service quality, and facility conditions, all of which aim to accurately show the differences in basic public services across regions. With the help of these indicators, we can more systematically analyze the situation of public services in various regions and provide a solid foundation for further data analysis and conclusions.

In addition, based on the principle of finding differences, these research indicators provide practical references for policy formulation and service improvement. Combining the previous research literature and taking into account the systematic nature of the indicator system, the availability of data and the representativeness of each indicator, this paper constructs an indicator system for evaluating the level of basic public services. This system is organized around seven first-level indicators, namely, education, public culture, medical treatment & public health, social security, employment services, environment and infrastructure services, which are further subdivided into 28 second-level indicators as shown in Table 1.

**Table 1. Evaluation system of basic public service level indicators**

| First-level indicators | Second-level indicators |
|---|---|
| Education | Per capita financial expenditure on education |
| | School enrollment ratio |
| | Pupil-teacher ratio of compulsory education period |
| | Student-teacher ratio of higher and vocational education period |
| | Patents granted per 10,000 people |
| | Inventions per 10,000 people |
| | Number of schools for compulsory education per 10,000 people |
| | Number of schools for higher and vocational education per 10,000 people |
| Public culture | Per capita access to library collections |
| | Number of museums per 10,000 people |
| Medical treatment & public health | Number of hospitals per 10,000 people |
| | Number of beds per 10,000 people |
| | Number of doctors per 10,000 people |
| Social security | Urban pension insurance coverage |
| | Urban medical insurance coverage |
| Employment service | Urban unemployment insurance coverage |
| | Labor market employment rate |
| | Average wage of employees |

| | Urban greening coverage |
|---|---|
| Environment | Non-hazardous treatment rate of domestic waste |
| | Centralized sewage treatment rate |
| | Annual average concentration of particulate matter |
| | Urban road space per 10,000 people |
| | Buses in operation per 10,000 people |
| Infrastructure | Public transportation coverage rate |
| | Gas consumption per 10,000 people |
| | Water consumption per 10,000 people |
| | Land for construction per 10,000 people |

## 4. Analysis of the current situation of the equalization level of basic public services in the Yangtze River Delta and its limiting factors

### 4.1 Status of basic public services in Yangtze River Delta based on coupling coordination degree model

#### 4.1.1 Construction of coupling coordination degree model

Here we introduce the concept of coupling coordination degree in physics (Li et al., 2022). Define the coupling coordination degree for $n$ subsystems is:

$$C_n = n \times \left[\frac{S_1 \times S_2 \times \ldots \times S_n}{\prod(S_i + S_j)}\right]^{\frac{1}{n}} \quad (2)$$

where $S_i$ denotes the contribution of the subsystem $i$ to the total system.

In this paper, only two subsystems are considered: the level of basic public services and the degree of modernization, so for the period of $t$, the coupling coordination model between the two subsystems of the city $i$ can be expressed as follows:

$$C_{2;i,t} = 2 \times \frac{\sqrt{S_{1;i,t} \times S_{2;i,t}}}{S_{1;i,t} + S_{2;i,t}} \quad (3)$$

Based on the degree of coupling that characterizes the relatedness of the subsystems, a degree of coordination model is then introduced to portray the developmental synergies between the subsystems in order to measure the degree of relatedness and coherent development between the systems more comprehensively. Therefore, the model is adapted to a coupling degree of coordination model:

$$D = \sqrt{C \times T}, \quad T = \alpha \times S_1 + (1 - \alpha) \times S_2 \quad (4)$$

where the contribution coefficient $\alpha$ is used to measure the importance of each subsystem in the whole system, $T$ is the comprehensive coordination index of subsystems, and $D$ is the degree of coupling coordination.

Since this paper studies the mutual coupling relationship between the level of basic public services and the degree of modernization, here we should assign the value of $\alpha$ to 0.5, indicating that the two systems are mutually reinforcing each other to the same degree of their relative importance.

Therefore, the degree of coupling coordination between the two subsystems of the city $i$ during the period $t$ can be expressed as:

$$D_{i,t} = \sqrt{C_{2;i,t} \times (0.5 \times S_{1;i,t} + 0.5 \times S_{2;i,t})} \qquad (5)$$

On the basis of existing researches, the types of coupled coordination between the basic public service system and the system of modernization degree are classified into 7 types and 4 stages as shown in Table 2.

**Table 2. Types of basic public service-modernization coupling coordination degree**

| Degree of coupled and coordinated development | Type of coupling | Coupling Stage |
| --- | --- | --- |
| $D \in [0,0.3)$ | severe disharmony and decline | low-level coupling |
| $D \in [0.3,0.4)$ | mild disharmony and decline | antagonism |
| $D \in [0.4,0.5)$ | close to disharmony and decline | |
| $D \in [0.5,0.6)$ | close to harmonious development | |
| $D \in [0.6,0.7)$ | moderate harmonious development | preliminary synergy |
| $D \in [0.7,0.8)$ | good harmonious development | |
| $D \in [0.8,1]$ | excellent harmonious development | high-level coupling |

**4.1.2 Analysis of findings and conclusions**

Based on the previous analysis and the existing evaluation system of basic public service indicators, this paper will then construct a systematic indicator evaluation system for the degree of modernization. To achieve this goal, the paper comprehensively selects economic development, industrial process and resource & environment as the first-level indicators, which are then subdivided into a total of 7 second-level indicators to comprehensively and multidimensionally measure the process and characteristics of regional modernization (Table 3).

**Table 3. Evaluation system for the level of development of regional modernization**

| First-level indicators | Second-level indicators |
| --- | --- |
| Economic development | Gross domestic product |
| | Level of import and export trade |
| Industrial progress | Level of development of the primary sector |
| | Level of development of the secondary industry |
| | Level of development of the tertiary industry |
| Resource & environment | Level of resource availability |
| | Level of environmental pollution |

Based on the data collected from the China Urban Statistical Yearbook, the CECN statistical database and other paths, and combined with the same data processing and indicator assignment methods as those used in previous sections, now we successfully obtained the quantitative results of the degree of modernization and development of each region. Further, the contribution of the two systems to the total system respectively can be calculated, so as to carry out the measurement of the coupling model.

For the results of the coupling model analysis, from the perspective of the overall time-series evolution, as shown in Figure 1, the annual average coupling coordination degree of each city in the past five years shows a basically stable and slightly upward trend, indicating that the degree of dual-system coupling has signs of enhancement. However, it drops sharply in 2019-2020. We speculate that maybe affected by the Covid-19 pandemic, the existing level of basic public service supply was insufficient to support the surge in demand, or the level of modernization experienced signs of reversal, leading to a slump in the degree of coupling coordinated development. Fortunately, its recovery accelerated in 2021, showing a positive trend in the degree of coupling.

According to the requirements of the regional modernization advancement on the development of basic public service equalization, it is considered that the level of basic public service equalization in each region is positively correlated with the degree of dual-system coupling development (same for the subsequent analysis in this section). Therefore, the curve of basic public service equalization level over time in the provinces and municipalities of the Yangtze River Delta also basically coincides with Figure 1, i.e., it shows a stable and positive trend over time, except for the sudden and sharp drop in 2020.

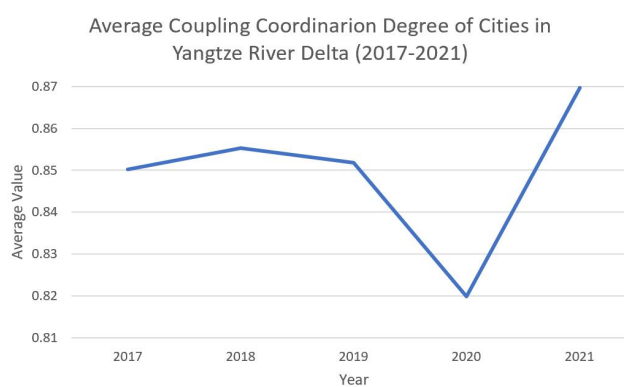

Figure 1. Average Coupling Coordination Degree of Cities in Yangtze River Delta (2017-2021)

Then we focus on the development characteristics of cities over time. As shown in Figure 2, during the period of 2017-2021, the coupling development degree of the vast majority of cities in YRD region is greater than 0.8. From the criteria in Table 2, it can be seen that the basic public services-modernization dual system of these cities belongs to the type of high-quality coordination, i.e., the development of basic public services and modernization are complementary, bringing positive impacts on the advancement paths of each other. Only in a very few cities are the two systems still in a period of preliminary synergy.

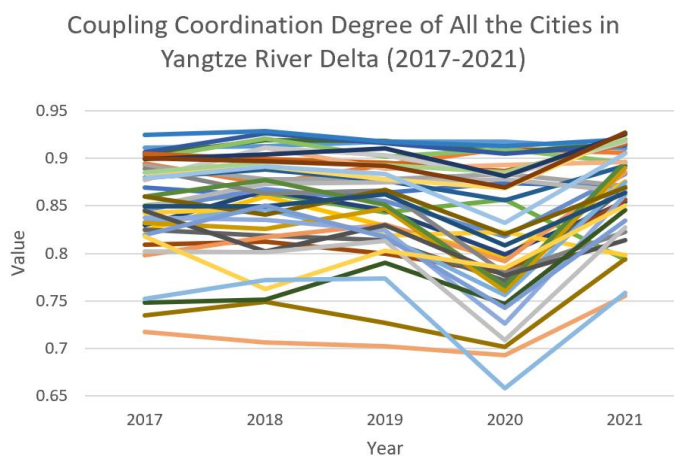

Figure 2. Coupling Coordination Degree of All the Cities in Yangtze River Delta (2017-2021)

Specifically, according to Table 4, there are five cities each year whose two systems were still in the stage of preliminary synergy in the period of 2017-2019. In 2020, 18 cities are characterized as preliminary synergy, while in 2021 this number comes back to 5. However, except for 2020, the degree of coupling and coordinated development shows an increasing trend, and the overall development of the two systems keeps improving in the long run.

In addition, we have to attach special attention to the cities of Yancheng, Fuyang, Lu'an and Xuancheng, which basically have not reached high level of coupling during the 5 years, indicating some specific shortcomings of their development need to be analyzed. In the following sections, this paper will try to deliver appropriate development suggestions for these cities.

Table 4. Coupling Coordination Degree of All the Cities in Yangtze River Delta Presented by Year (2017-2021)

|  | 2017 | 2018 | 2019 | 2020 | 2021 |
| --- | --- | --- | --- | --- | --- |
| Shanghai | 0.8692 | 0.8607 | 0.8634 | 0.8738 | 0.8704 |
| Nanjing | 0.8944 | 0.8735 | 0.8926 | 0.9104 | 0.9054 |
| Wuxi | 0.8924 | 0.8777 | 0.8780 | 0.8836 | 0.8708 |
| Xuzhou | 0.8243 | 0.8597 | 0.8286 | 0.7962 | 0.8756 |
| Changzhou | 0.9112 | 0.9131 | 0.9176 | 0.9173 | 0.9103 |
| Suzhou | 0.8505 | 0.8641 | 0.8431 | 0.8564 | 0.7938 |
| Nantong | 0.8294 | 0.8662 | 0.8452 | 0.7988 | 0.8539 |
| Lianyungang | 0.8089 | 0.8124 | 0.7994 | 0.7795 | 0.8568 |
| Huai'an | 0.8248 | 0.8189 | 0.8136 | 0.7711 | 0.8682 |
| Yancheng | 0.7348 | 0.7487 | 0.7270 | 0.7021 | 0.7941 |
| Yangzhou | 0.8797 | 0.8881 | 0.8762 | 0.8555 | 0.8926 |
| Zhenjiang | 0.9035 | 0.9189 | 0.9181 | 0.9067 | 0.9153 |
| Taizhou | 0.8459 | 0.8673 | 0.8547 | 0.8179 | 0.8832 |
| Suqian | 0.7980 | 0.8173 | 0.8305 | 0.7917 | 0.8849 |
| Hangzhou | 0.8486 | 0.8728 | 0.8748 | 0.8779 | 0.8653 |
| Ningbo | 0.8424 | 0.8483 | 0.8184 | 0.8227 | 0.7982 |
| Wenzhou | 0.8326 | 0.8463 | 0.8177 | 0.7563 | 0.8751 |
| Jiaxing | 0.8990 | 0.9210 | 0.9018 | 0.9098 | 0.8954 |
| Huzhou | 0.9063 | 0.9265 | 0.9157 | 0.9047 | 0.9133 |
| Shaoxing | 0.9047 | 0.8996 | 0.8950 | 0.8864 | 0.9149 |
| Jinhua | 0.8893 | 0.8630 | 0.8664 | 0.7795 | 0.8229 |
| Quzhou | 0.8319 | 0.8261 | 0.8471 | 0.7604 | 0.8896 |
| Zhoushan | 0.9248 | 0.9284 | 0.9172 | 0.9124 | 0.9196 |
| Taizhou | 0.8599 | 0.8769 | 0.8514 | 0.7658 | 0.8960 |
| Lishui | 0.8379 | 0.8354 | 0.8241 | 0.7260 | 0.8614 |
| Hefei | 0.8772 | 0.9110 | 0.8888 | 0.8930 | 0.8959 |
| Wuhu | 0.8787 | 0.9109 | 0.9035 | 0.8813 | 0.9172 |
| Bengbu | 0.8851 | 0.8915 | 0.8778 | 0.8684 | 0.9238 |
| Huainan | 0.8789 | 0.8911 | 0.8838 | 0.8318 | 0.9046 |
| Ma'anshan | 0.8851 | 0.8961 | 0.8934 | 0.8852 | 0.9189 |
| Huaibei | 0.8992 | 0.9040 | 0.9102 | 0.8808 | 0.9257 |
| Tongling | 0.9000 | 0.8965 | 0.8922 | 0.8695 | 0.9271 |
| Anqing | 0.8446 | 0.8018 | 0.8294 | 0.7767 | 0.8139 |
| Huangshan | 0.8594 | 0.8405 | 0.8666 | 0.8206 | 0.8691 |
| Chuzhou | 0.8492 | 0.8497 | 0.8619 | 0.8081 | 0.8641 |
| Fuyang | 0.7482 | 0.7513 | 0.7902 | 0.7468 | 0.8458 |
| Suzhou* | 0.8191 | 0.8505 | 0.8172 | 0.7429 | 0.8348 |
| Lu'an | 0.7176 | 0.7063 | 0.7028 | 0.6928 | 0.7555 |
| Bozhou | 0.8019 | 0.8012 | 0.8131 | 0.7089 | 0.8276 |
| Chizhou | 0.8176 | 0.7623 | 0.8025 | 0.7844 | 0.8501 |
| Xuancheng | 0.7521 | 0.7719 | 0.7736 | 0.6585 | 0.7583 |

Note: Suzhou* is referred to Suzhou City in Anhui Province, not the one in Jiangsu Province which shares the same spelling. If mentioned, Suzhou City in Jiangsu Province will be shown without a star throughout this paper.

## 4.2 Analysis of equalization of basic public service level in Yangtze River Delta based on Moran's I index

Moran's index is divided into global Moran's I index and local Moran's I index, which can be used to analyze the spatial autocorrelation characteristics and measures the degree of spatial aggregation. The global Moran's I index represents the global autocorrelation and the local Moran's I index measures the local autocorrelation.

### 4.2.1 Analysis of the overall equalization level of development by the global spatial Moran's I index

First we give the explanation of the global Moran's I index equation:

$$I = \frac{n \sum_{i=1}^{n} \sum_{j=1}^{n} W_{ij}(y_i - \bar{y})(y_j - \bar{y})}{\sum_{i=1}^{n} \sum_{j=1}^{n} W_{ij} \sum_{i=1}^{n}(y_i - \bar{y})^2} \quad (6)$$

In this equation, $I$ is defined as Moran's I index, $n$ is the number of study units, $W$ is the spatial weight matrix, where neighboring units take the value of 1 and others take the value of 0. Meanwhile, $y_i$ and $y_j$ denote the values of public service equalization level of province or city $i$ and province or city $j$ respectively, $\bar{y}$ denotes the average value of public service equalization level.

Using the data obtained in the previous public service equalization evaluation system, the global Moran's I was solved by MATLAB software and the results were obtained as shown in Table 5:

**Table 5. Global Moran's I Index (2017-2021)**

| Year | Moran's I index | z-score | p-value |
|---|---|---|---|
| 2017 | 0.273813 | 2.478996 | 0.006588 |
| 2018 | 0.253848 | 2.305177 | 0.010578 |
| 2019 | 0.267807 | 2.81883 | 0.00241 |
| 2020 | 0.253014 | 2.683069 | 0.003647 |
| 2021 | 0.226435 | 2.424135 | 0.007672 |

Note: p-value denotes significance and z-score denotes the standard deviation multiplier, which captures the degree of dispersion of a dataset.

From the results in the table, the p-values from 2017 to 2021 are all less than 0.05,

passing the significance test. Meanwhile, the global Moran's I value of each city in the Yangtze River Delta region is all greater than 0 during the 5 years, which indicates a generally positive correlation between cities, meaning that aggregation is more likely to occur in places where the level of equalization of basic public services is higher or lower. At the same time, we notice that the Moran's I index shows an overall decreasing trend from 2017 to 2021, showing a decreasing trend of the degree of aggregation.

**4.2.2 Autocorrelation measured by the local spatial Moran's I index**

Initially, inspired by the application of Moran's I index in existing researches (Chen, 2013), we'll use the LISA aggregation map obtained from the local Moran's I index to analyze the degree of spatial correlation between each region and its neighbors. According to the value of its Moran's I index, the regions can be divided into four types, which are high-high aggregation (HH), high-low aggregation (HL), low-low aggregation (LL), and low-high aggregation (LH) respectively.

For the regional unit of $i$, Moran's I index LISA equation is:

$$I_i = \frac{y_i - \bar{y}}{S^2} \sum_{i,j=1}^{n} W_{ij}(y_j - \bar{y}),$$

$$S^2 = \frac{1}{n}\sum_{i=1}^{n}(y_i - \bar{y})^2 \tag{7}$$

where $i \neq j$, $n$ is the number of space units, $y_i$ and $y_j$ denote the observation of $x$ on regions $i$ and $j$ for a certain situation respectively, and $W$ is the spatial weight matrix (Zhang et al., 2022).

After the obtained indices were aggregated, we then use the GeoDa software to generate LISA aggregation maps of the YRD region about the equalization level of basic public services for each year from 2017 to 2021, shown as following Figure 3-1 to Figure 3-5.

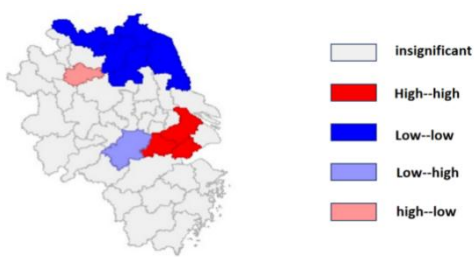

Figure 3-1. LISA Aggregation Map (2017)

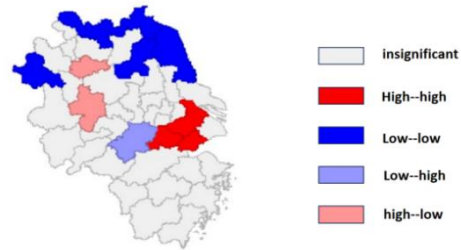

Figure 3-2. LISA Aggregation Map (2018)

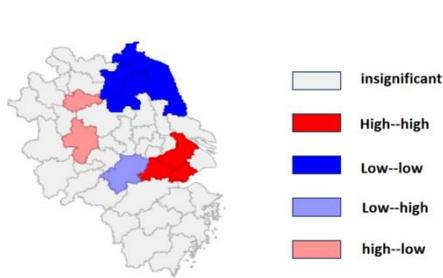 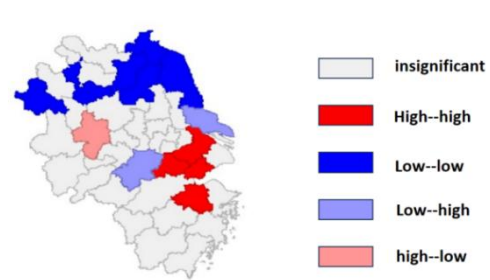

Figure 3-3. LISA Aggregation Map (2019)  |  Figure 3-4. LISA Aggregation Map (2020)

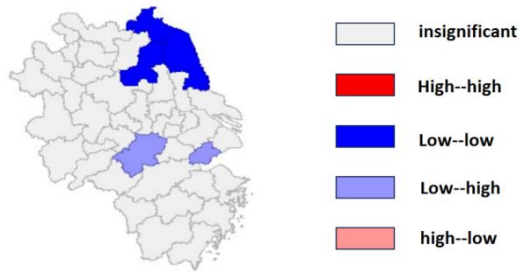

Figure 3-5. LISA Aggregation Map (2021)

From the LISA maps, it's obvious that the 4 cases of high-high, high-low, low-low and low-high aggregation of the level of basic public service equalization can all be found in the YRD region during the years of 2017-2021.

**High-high aggregation situation**: From 2017 to 2019, the high-high aggregation mainly appears in the cities of Suzhou, Huzhou, and Jiaxing. In 2020, the city of Shaoxing starts to show high-high aggregation. And no high-high aggregation areas appear in 2021.

**Low-low aggregation situation**: From 2017 to 2021, the low-low aggregation mainly concentrates in the northeastern region of the Yangtze River Delta, while the cities of Lianyungang, Yancheng, and Huai'an are all low-low aggregation areas during this five-year period. The cities of Suqian and Xuzhou both show low-low aggregation in 2017. However, the situation for Xuzhou only lasts one year, while Suqian remains as a low-low aggregation area in 2019 and 2020. Meanwhile, in 2020, Fuyang, Huaibei and Bengbu are also areas with low-low aggregation.

**Low-high aggregation situation**: From 2017 to 2021, Xuancheng remains as a low-high aggregation area. Nantong and Jiaxing show low-high aggregation in the year of 2020 and 2021 respectively.

**High-low aggregation situation**: Bengbu City has high-low aggregation zones from 2017 to 2019, and Hefei City has high-low aggregation zones from 2018 to 2020. There are no high-low aggregation zones in 2021.

**Analytical summary**: In recent years, the low-low aggregation areas gradually decrease, indicating that the level of equalization of basic public services has increased in the northeast region of the Yangtze River Delta, except for a tendency to increase in 2020 which is presumed to be probably due to the reduction of inputs of basic public services caused by the epidemic. Meanwhile, the phenomenon of high-high and high-low aggregation zones disappeared in 2021, proving that the imbalance in the development of basic public services in the Yangtze River Delta region has been reduced.

## 4.3 Analysis of the main causes of spatial variability in the level of basic public services based on the Theil Index

The measurement of regional agglomeration based on the level of basic public services finds that the spatial correlation is generally strengthening the synergistical development of cities in the YRD region. The following analyses focus on the measurement of specific spatial variability, where we introduce the concept of Theil index (Shen et al., 2023).

### 4.3.1 Theil index measurement methodology

Divide the $m$ cities into $k$ groups, each group is denoted by $G_k$ (k=1,2,3,4), the number of cities in the $No.k$ group is set as $m_k$. Denote the proportion of the level of equalization of basic public services of the $No.i$ city to the overall by $x_i$, and then the Theil index is:

$$T = \sum_{k=1}^{4} p_k \ln \frac{p_k}{m_k/m} \qquad (8)$$

Denote the level of equalization of urban public services in group $k$ as a share of the total by $p_k$. Let $T_b$ and $T_w$ represent between-group and within-group differences respectively, then we can decompose the Thiel index as:

$$\sum_{k=1}^{4} m_k = m$$

$$T = T_b + T_w = \sum_{k=1}^{4} p_k \ln \frac{p_k}{m_k/m} + \sum_{k=1}^{4} p_k T_k$$

$$T_k = \sum_{i \in G_k} \frac{x_i}{p_k} \ln \frac{x_i/p_k}{1/m_k} \qquad (9)$$

$T_k$ is the within-group difference for group $k$. The value of the Theil index ranges from 0 to 1. When it takes the value of 1, it indicates complete inequality, and when it is 0, it indicates complete equality in the level of public services in the region.

Further, the contribution of intra-group gaps and the contribution of inter-group gaps can be calculated for group $k$:

$$D_k = p_k \frac{T_k}{T}, \quad D_b = \frac{T_b}{T} \qquad (10)$$

### 4.3.2 Analysis of the results of the Theil Index

We use three provinces (Jiangsu, Zhejiang and Anhui) and one city (Shanghai) as the criteria for grouping to calculate the Theil index of the level of equalization of basic public services in the YRD region.

**Table 6. Theil Index of Equalization Level of Basic Public Services in YRD Region Presented by Year (2017-2021)**

| Year | Population discrepancy | Differences within groups | | | | | | Differences between groups | |
|---|---|---|---|---|---|---|---|---|---|
| | | Shanghai | Jiangsu Province | Zhejiang Province | Anhui Province | Contribution value | Contribution (%) | Contribution Value | Contribution (%) |
| 2017 | 0.3633 | 0 | 0.2996 | 0.0709 | 0.0420 | 0.2358 | 0.7051 | 0.1275 | 0.2949 |
| 2018 | 0.3356 | 0 | 0.3695 | 0.1774 | 0.1203 | 0.2176 | 0.6484 | 0.11801 | 0.3516 |
| 2019 | 0.1680 | 0 | 0.2127 | 0.0947 | 0.0427 | 0.1142 | 0.6797 | 0.05381 | 0.3203 |
| 2020 | 0.1930 | 0 | 0.2432 | 0.0998 | 0.0507 | 0.1284 | 0.6651 | 0.06464 | 0.3349 |
| 2021 | 0.1564 | 0 | 0.2194 | 0.0889 | 0.0393 | 0.1123 | 0.7184 | 0.04404 | 0.2816 |

The results of Table 6 show that the level of equalization of basic public services in the YRD region exhibits the following characteristics:

**a.** Generally, there is a downward trend in the Theil Index of the level of equalization of basic public services in the Yangtze River Delta region, which indicates that there is an overall downward trend in the regional differences in the level of basic public services in the Yangtze River Delta region, especially in 2019, when there is a larger decline. This may be due to the fact that the Outline of the Integrated Regional Development of the Yangtze River Delta (State Council, 2019) was officially released and implemented in 2019, so the decline in regional differences is in line with the expectation of integrated development. The overall difference between 2017 and 2018 is greater than 0.3, suggesting that the level of equalization of basic public services in the three provinces and one city still varied considerably before the implementation of specific policies, which might be the exact target expected to be improved through regional integration. After 2019, the Theil indices fall below 0.2 in all cases, indicating that the differences have been relatively smaller.

**b.** After decomposition, the analysis of Theil index shows that the contribution rates of both intra-group differences and inter-group differences are relatively stable. The contribution rate of intra-group differences is larger, with its proportion reaching about 70%, while the contribution rate of inter-group differences is around 30%. The decomposition results shows that intra-group differences are the main source of overall differences. Among them, Jiangsu Province is confronting larger intra-group differences, while the differences within groups faced by Anhui Province is relatively smaller, indicating that Jiangsu Province should pay more attention to the problem of intra-group differences in equalization of basic public services and try to improve current situation. At the same time, we can find that the provision of basic public services in Zhejiang Province is more balanced, in accordance with its relatively higher level of equalization during the same period, which may partially explain why Zhejiang Province has been entrusted with the historical mission of high-quality development to construct a model area of common prosperity (State Council, 2021).

Further, the level of equalization of basic public services in the Yangtze River Delta region in 2021 is analyzed and measured by dimension.

**Table 7. Theil Index of Equalization Level of Basic Public Services in YRD Region Presented by Dimension (2021)**

| Dimension | Population discrepancy | Differences within groups | | | | | | Differences between groups | |
|---|---|---|---|---|---|---|---|---|---|
| | | Shanghai | Jiangsu Province | Zhejiang Province | Anhui Province | Contribution value | Contribution (%) | Contribution Value | Contribution (%) |
| Education | 0.0943 | 0 | 0.4748 | 0.1505 | 0.1690 | 0.0749 | 79.43% | 0.0194 | 20.57% |
| Public culture | 0.4132 | 0 | 0.2261 | 0.1117 | 0.2457 | 0.2411 | 58.35% | 0.1721 | 41.65% |

| | | | | | | | | |
|---|---|---|---|---|---|---|---|---|
| Medical treatment & public health | 0.2338 | 0 | 0.3033 | 0.2587 | 0.3724 | 0.2185 | 93.45% | 0.0153 | 6.55% |
| Social security | 0.3485 | 0 | 0.2682 | 0.1044 | 0.1002 | 0.1648 | 47.28% | 0.1837 | 52.72% |
| Employment service | 0.3970 | 0 | 0.3141 | 0.1239 | 0.0874 | 0.2086 | 52.54% | 0.1884 | 47.46% |
| Environment | 0.0180 | 0 | 0.3111 | 0.1674 | 0.2654 | 0.0134 | 74.39% | 0.0046 | 25.61% |
| Infrastructure | 0.4725 | 0 | 0.1937 | 0.3093 | 0.2869 | 0.3733 | 79.00% | 0.0993 | 21.00% |

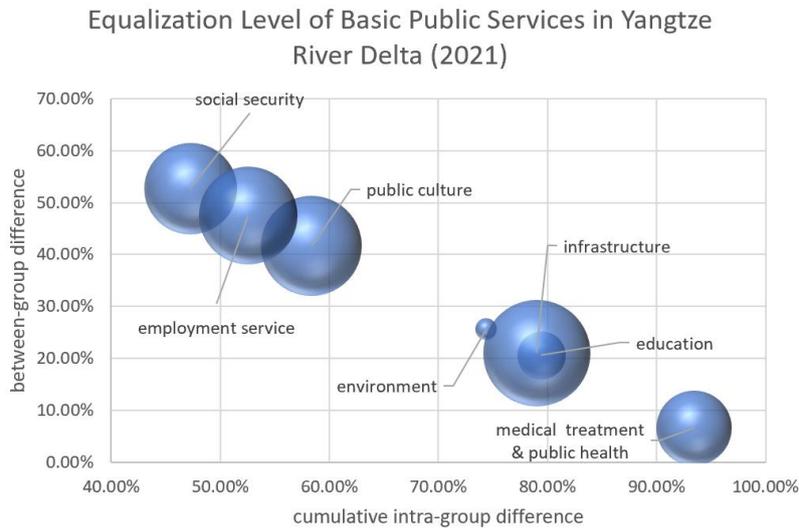

Figure 4. Equalization Level of Basic Public Services in YRD Region (2021)

As can be seen from Table 7 and Figure 4, among all the 7 dimensions, infrastructure services exhibit the greatest differences, followed by social security services, employment services, public culture services and medical treatment & public health services. The levels of non-equalization in education services and environmental services are decreasing but non-significant. Additionally, the intra-group differences in social security services, employment services and public culture services are relatively balanced with the inter-group differences, while the intra-group differences in medical treatment & health services are particularly prominent compared with those between groups, indicating the particularly imbalanced level of supply in medical and health services at the county level within each province and city, which is in line with previous scholars' emphasis on enhancing urban healthcare service accessibility to build more sustainable and inclusive societies (Xia et al., 2022).

Overall, the differences between groups are not particularly striking, while the differences within groups turn out to be more significant, reminding us that it is the intra-provincial and municipal differences that need to be first eliminated in the current push to enhance the level of equalization of basic public services.

## 5. Exploration of the Development Path towards Equalization of Basic Public Services in the Yangtze River Delta

### 5.1 AF model for identifying multidimensional shortcomings from the imbalance of basic public services

Previous calculations indicate that the level of equalization of public services in some cities is weak, and further exploration obtained the conclusion that there is a significant difference between cities within group. In order to dig deeper into the basic public service shortcomings of each city, this paper adopts the AF method to further identify the weak counties and their shortcomings, and then focuses on the weak counties above to conduct a detailed analysis of the grading indicators, in order to putting forward the county-level targeted policy recommendations to promote the equalization of basic public services. In this case, based on the bottom 10 cities selected from the five-year average scores of each city (Figure 5), this paper will then mainly analyze the 40 counties under the jurisdiction of these cities in the following sections.

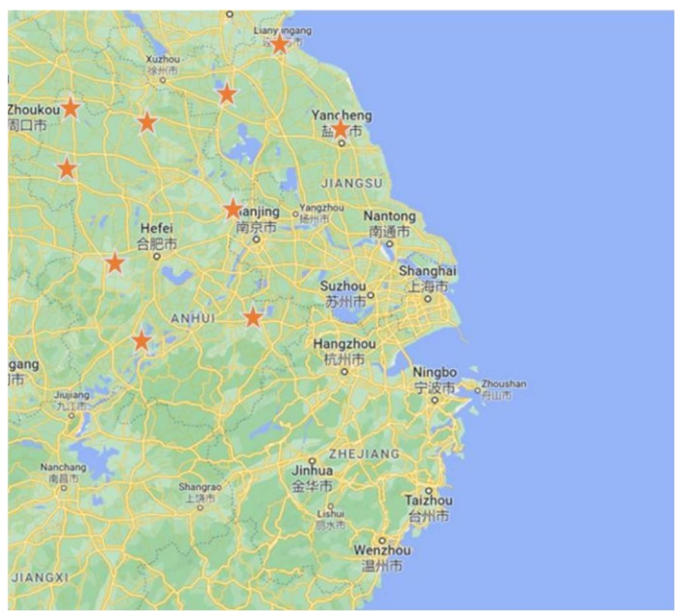

Figure 5. Schematic Diagram of the Bottom 10 cities in Terms of the Level of Basic Public Services (highlighted by the symbol of star)

### 5.1.1 AF modeling basic principles and steps

**a.** Construct $m \times n$ dimensional county public service level matrix $A$, where $m$ represents the number of counties to be analyzed, and $n$ represents the number of selected public service indicators, i.e. $A_{ij}$ represents the value of the $No.j$ public service indicator for county $i$.

**b.** Distinguish public service shortcomings. Construct $n$ dimensional shortboard threshold vector $X$, where each component represents the shortboard threshold value of the corresponding public service level, and then construct an $m \times n$ dimensional county public service shortboard 0-1 matrix $B$, letting $B_{ij}$ represent the county $i$ in the first $j$ shortage judgment value of the public service indicator. When $A_{ij} < X_j$, let judgement value $B_{ij}$ equal 1, representing the existence of short board, and vice versa, granting $B_{ij}$ the value of 0 represents the non-existence of short board.

**c.** Set $n$ dimensional column vector $Y$ to represent the weight of a county's basic public service short board for measurement, construct the identification mapping $f: R^n \to \{0,1\}$, take each row vector $B_i$ of matrix $B$, if $B_i Y > Q$, then if $f$ takes 1, it means the county has public service weakness problem, and vice versa, taking $f$ as 0 is recorded as non-public service weakness in the county. And the mapping above is not only related to the multidimensional shortages, but also affected by the shortages of each indicator, so it is called double criticality method.

**d.** Define the count correction matrix $C_{ij}$, for the above mapping $f$, if it is taken to be 0, then replace the rows of the counties corresponding to the original shortcoming matrix with the $n$-dimension **0** vector. Define the count correction vector **Z**, where each component is the dot product of the row vector of the count correction matrix and the weight vector **Y**, which represents the weighted aggregation of the number of dimensions in which there are shortcomings in counties with weak public services.

**e.** Calculate the public service shortcoming index. Define the $n$ dimensional vector **W** to represent the public service shortages exclusion, where $W_i = f_i$, define the incidence of public service shortages as $U = \sum_{i=1}^{m} \frac{W_i}{m}$, the degree of shortage of public service short board as $T = \frac{\sum_{i=1}^{m} C_i Y}{\sum_{i=1}^{m} W_i}$.

**f.** Define the public service shortcoming index as $M = U \times T$. The index can then be disaggregated by indicator or by region, granted by its excellent disaggregation properties to reflect the extent to which each indicator and subregion affects the index.

**5.1.2 Selection of AF model indicators**

Through the combination of entropy weighting and coefficient of variation weighting method, it is found that the weighting of indicators such as the student-teacher ratio of higher and vocational education period is very low, which displays its little influence on the degree of public service equalization. Based on previous research and field investigation, this paper introduces the dimension of industrial economy, combining with modernization, to re-select the indicators and construct the measurement scheme of the shortboards of public services shown as Table 8. From constructing the public service shortboard index, we then continue to analyze the short board problems in these counties' basic public services, in an attempt to provide targeted policy suggestions.

**Table 8. Evaluation System of Indicators for Measuring Shortcomings in Basic Public Services**

| First-level indicators | Second-level indicators |
| --- | --- |
| Industrial services | Digital economy index |
| | Participation in the cooperative economy |
| | Share of agricultural mechanization |
| Cultural services | Number of museums per 10,000 people |
| | Number of cultural centers per 10,000 people |
| | Number of books per 10,000 people |
| Ecological services | Urban greening coverage |
| | Garbage disposal rate |
| | Sewage treatment rate |
| Medical services | Hospitals per 10,000 people |
| | Beds per 10,000 people |

| | Life services | Penetration rate of water supply |
| --- | --- | --- |
| | | Public transportation & road density |

Then the MPI module in Stata was used to program and calculate the results.

### 5.1.3 Results and analysis of the public service shortcoming index

Based on previous researches on big data analysis and empirical studies, this paper selects $k = 0.5$ as the criterion for identifying weak counties in basic public services. The results of the shortboard index are as follows:

**Table 9. Results of the Analysis of the Shortcomings Index**

| Threshold value of shortboards | Incidence of shortboards (U) | Number of counties with weak public services | Shortboard level (T) | Shortboard index (M) |
| --- | --- | --- | --- | --- |
| 0.8 | 0.064 | 3 | 0.932 | 0.0599 |
| 0.75 | 0.100 | 4 | 0.856 | 0.0853 |
| 0.7 | 0.118 | 5 | 0.811 | 0.0957 |
| 0.65 | 0.165 | 7 | 0.789 | 0.1302 |
| 0.6 | 0.212 | 8 | 0.751 | 0.1592 |
| 0.55 | 0.310 | 12 | 0.691 | 0.2142 |
| 0.5 | 0.321 | 13 | 0.612 | 0.1965 |
| 0.45 | 0.439 | 18 | 0.558 | 0.2450 |
| 0.4 | 0.541 | 22 | 0.521 | 0.2819 |
| 0.35 | 0.591 | 24 | 0.441 | 0.2606 |
| 0.3 | 0.665 | 27 | 0.423 | 0.2813 |
| 0.25 | 0.763 | 31 | 0.394 | 0.3006 |
| 0.2 | 0.821 | 33 | 0.372 | 0.3054 |
| 0.15 | 0.896 | 36 | 0.355 | 0.3181 |
| 0.1 | 0.943 | 38 | 0.324 | 0.3055 |

As shown in Table 9, a total of 13 counties with weak public service problems are identified, and the analysis above also reveals an average of 4.8 indicators of shortcomings per county with weak public services. Through further calculation, we then find that the multidimensional public service shortcoming index for the 10 city's 40 counties in 2021 is 0.196, and the proportion of counties with problems of weak public services in the 10-city area is 32.5%.

This paper then focuses on the analysis of inter-city differences in the counties above. The results of the 2017 and 2021 municipal public service weakness index measurements are shown in Figure 6.

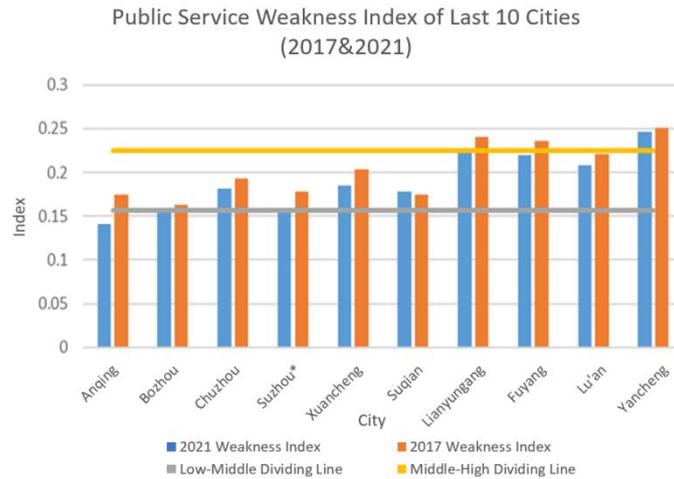

Figure 6. Public Service Weakness Index of Last 10 Cities in YRD Region (2017&2021)

From the latest measurement of 2021, the city with the lowest shortfall index is: Anqing, while the city with the highest shortfall index is: Yancheng. The mean value of the public service shortcoming index for the 10 cities' 40 counties is 0.191, and the standard deviation of the public service shortcoming index for the counties analyzed is 0.034, thus indicating that there are large differences in the public service shortcoming indices between different cities.

In addition, according to the relationship between the mean and standard deviation of the shortboard index mentioned above, the situations different cities confronts can be divided into three types: low shortboard (score lower than the mean of the shortboard index minus the variance), medium shortboard (score between the mean of the shortboard index minus the variance and plus the variance), and high shortboard (score higher than the mean of the shortboard index plus the variance), and it is clear from Figure 6 to figure out Anqing, Huizhou and Suzhou* City all exhibit low-level shortness, while Chuzhou, Xuancheng, Suqian, Lu'an and Fuyang City all present medium shortness, with Lianyungang & Yancheng City confronted with high-level shortness.

By comparing the county public service shortfall index of each city in 2021 with that of 2017, we notice that the shortcoming level of most cities in the YRD region has declined to a certain extent, especially Anqing and Suzhou*, but there are still significant differences between different cities.

Finally, the shortcoming index and the contribution of each indicator corresponding to the counties are analyzed and shown in Table 10.

**Table 10. The Shortcoming Index of Basic Public Services of the 10 Weakest Cities' 40 Counties in YRD Region**

|  | Ocr | deg | *Ind* | A | B | C | D | E | F | G | H | I | J | K | L | M |
|---|---|---|---|---|---|---|---|---|---|---|---|---|---|---|---|---|
| Dingyuan | 0.64 | 0.64 | 0.41 | 0.03 | 0.04 | 0.02 | 0.02 | 0.05 | 0.03 | 0.01 | 0.03 | 0.02 | 0.05 | 0.03 | 0.01 | 0.03 |
| Jixi | 0.49 | 0.74 | 0.36 | 0.04 | 0.03 | 0.02 | 0.02 | 0.04 | 0.02 | 0.04 | 0 | 0 | 0.03 | 0 | 0.03 | 0.04 |
| Jinzhai | 0.87 | 0.42 | 0.32 | 0.02 | 0.03 | 0.04 | 0.02 | 0.05 | 0.03 | 0.03 | 0.04 | 0.02 | 0.05 | 0 | 0 | 0.01 |
| Susong | 0.4 | 0.77 | 0.31 | 0.01 | 0.02 | 0.02 | 0.01 | 0.04 | 0.03 | 0.01 | 0.01 | 0.01 | 0.01 | 0 | 0.03 | 0.02 |
| Guannan | 0.43 | 0.69 | 0.3 | 0.02 | 0.01 | 0.04 | 0.01 | 0 | 0.01 | 0.01 | 0 | 0.05 | 0.02 | 0.04 | 0.02 | 0.02 |
| Yingshang | 0.34 | 0.86 | 0.29 | 0.04 | 0.04 | 0.03 | 0.04 | 0 | 0.03 | 0.02 | 0.03 | 0.03 | 0.05 | 0.02 | 0.02 | 0.02 |

| | Ocr | deg | Ind | A | B | C | D | E | F | G | H | I | | | | |
|---|---|---|---|---|---|---|---|---|---|---|---|---|---|---|---|---|
| Sheyang | 0.66 | 0.44 | 0.29 | 0.04 | 0.04 | 0.03 | 0.04 | 0.02 | 0.03 | 0 | 0.02 | 0.04 | 0.04 | 0.02 | 0.05 | 0.04 |
| Siyang | 0.57 | 0.48 | 0.27 | 0.04 | 0.02 | 0.02 | 0.03 | 0.04 | 0.01 | 0.02 | 0.04 | 0.01 | 0.01 | 0.02 | 0.04 | 0.04 |
| Sihong | 0.64 | 0.43 | 0.27 | 0.03 | 0.03 | 0 | 0.03 | 0.03 | 0.01 | 0.02 | 0.02 | 0.03 | 0.02 | 0.01 | 0.03 | 0.01 |
| Dangshan | 0.34 | 0.78 | 0.27 | 0.02 | 0.01 | 0.03 | 0.01 | 0.03 | 0.01 | 0.03 | 0.04 | 0.05 | 0.02 | 0.03 | 0.04 | 0.03 |
| Binhai | 0.35 | 0.77 | 0.27 | 0.04 | 0.02 | 0.04 | 0.05 | 0.02 | 0.02 | 0.01 | 0.04 | 0.02 | 0.03 | 0 | 0.01 | 0.03 |
| Guanyun | 0.38 | 0.68 | 0.26 | 0.01 | 0.02 | 0.01 | 0.01 | 0.01 | 0.01 | 0.03 | 0.01 | 0.01 | 0.01 | 0.05 | 0.04 | 0.02 |
| Shuyang | 0.29 | 0.85 | 0.25 | 0.02 | 0.05 | 0.03 | 0.05 | 0.03 | 0.01 | 0.03 | 0.05 | 0.02 | 0.03 | 0.03 | 0.02 | 0.04 |
| Quanjiao | 0.59 | 0.42 | 0.25 | 0.03 | 0.01 | 0.04 | 0.01 | 0.04 | 0.01 | 0.01 | 0.04 | 0.02 | 0.03 | 0.02 | 0.01 | 0.02 |
| Langxi | 0.42 | 0.6 | 0.25 | 0.04 | 0.01 | 0.02 | 0.01 | 0.02 | 0.02 | 0.05 | 0.02 | 0.02 | 0.03 | 0.03 | 0 | 0.03 |
| Xiangshui | 0.3 | 0.77 | 0.24 | 0.04 | 0.02 | 0.02 | 0.01 | 0.04 | 0.03 | 0 | 0 | 0.01 | 0 | 0.02 | 0.01 | 0.03 |
| Jianhu | 0.53 | 0.45 | 0.24 | 0.03 | 0.01 | 0.05 | 0.05 | 0.01 | 0.02 | 0.04 | 0.04 | 0 | 0.01 | 0.04 | 0.02 | 0.01 |
| Linquan | 0.25 | 0.9 | 0.23 | 0.04 | 0.03 | 0 | 0.05 | 0.02 | 0.02 | 0.03 | 0.01 | 0.02 | 0.02 | 0.03 | 0 | 0.03 |
| Yuexi | 0.23 | 0.86 | 0.2 | 0.03 | 0.01 | 0.03 | 0.02 | 0.03 | 0.02 | 0.04 | 0.02 | 0.01 | 0.02 | 0.01 | 0.04 | 0.01 |
| Lixin | 0.34 | 0.54 | 0.19 | 0 | 0.05 | 0.01 | 0.03 | 0.02 | 0.01 | 0.03 | 0.04 | 0.04 | 0.03 | 0.01 | 0 | 0.02 |
| Huoshan | 0.83 | 0.21 | 0.18 | 0.01 | 0.02 | 0.04 | 0 | 0.05 | 0.05 | 0.02 | 0.03 | 0.03 | 0.02 | 0.03 | 0.04 | 0.01 |
| Taihu | 0.17 | 0.88 | 0.17 | 0.04 | 0.03 | 0 | 0.02 | 0.04 | 0.02 | 0.03 | 0 | 0.05 | 0.01 | 0.04 | 0.02 | 0.01 |
| Taihe | 0.43 | 0.4 | 0.17 | 0.04 | 0.03 | 0.03 | 0.01 | 0.04 | 0.03 | 0.02 | 0.04 | 0 | 0.01 | 0.02 | 0.03 | 0.04 |
| Lai'an | 0.26 | 0.64 | 0.17 | 0.01 | 0.01 | 0.04 | 0.01 | 0 | 0.02 | 0.03 | 0.01 | 0.05 | 0.01 | 0.04 | 0.05 | 0.03 |
| Funan | 0.6 | 0.9 | 0.17 | 0.01 | 0.01 | 0 | 0.02 | 0.01 | 0.01 | 0.05 | 0.01 | 0.01 | 0.03 | 0.03 | 0.05 | 0.02 |
| Guoyang | 0.58 | 0.27 | 0.16 | 0.01 | 0.04 | 0.01 | 0.05 | 0.03 | 0.03 | 0.02 | 0.05 | 0.03 | 0.04 | 0.03 | 0.01 | 0.03 |
| Wangjiang | 0.32 | 0.5 | 0.16 | 0.01 | 0.01 | 0.05 | 0.05 | 0.03 | 0.02 | 0.02 | 0.04 | 0.01 | 0.02 | 0.03 | 0.03 | 0.01 |
| Si | 0.65 | 0.25 | 0.16 | 0.03 | 0.05 | 0.03 | 0 | 0.01 | 0.05 | 0.01 | 0 | 0.02 | 0.05 | 0.01 | 0.03 | 0.01 |
| Mengcheng | 0.6 | 0.27 | 0.16 | 0 | 0.01 | 0.05 | 0.05 | 0.02 | 0.04 | 0.04 | 0.03 | 0.03 | 0.01 | 0.05 | 0.02 | 0.03 |
| Huaining | 0.2 | 0.79 | 0.16 | 0.04 | 0.02 | 0.04 | 0.03 | 0.02 | 0.01 | 0.04 | 0.02 | 0.03 | 0 | 0.03 | 0.03 | 0.05 |
| Fengyang | 0.29 | 0.55 | 0.16 | 0 | 0.01 | 0 | 0 | 0.04 | 0.01 | 0.03 | 0.03 | 0.04 | 0.05 | 0.05 | 0.04 | 0.03 |
| Xiao | 0.16 | 0.99 | 0.15 | 0.03 | 0.04 | 0.01 | 0.02 | 0.04 | 0.05 | 0 | 0.03 | 0.01 | 0.04 | 0.01 | 0.02 | 0.01 |
| Shucheng | 0.19 | 0.77 | 0.15 | 0.04 | 0.04 | 0.02 | 0.01 | 0.03 | 0.03 | 0.01 | 0.04 | 0.03 | 0.01 | 0 | 0.03 | 0.03 |
| Guangde | 0.25 | 0.61 | 0.15 | 0.03 | 0.04 | 0.01 | 0.05 | 0.03 | 0.02 | 0 | 0 | 0.05 | 0.01 | 0.04 | 0.01 | 0.03 |
| Funing | 0.27 | 0.55 | 0.15 | 0.01 | 0.04 | 0.03 | 0 | 0.01 | 0.04 | 0.04 | 0.02 | 0.04 | 0.05 | 0.03 | 0.04 | 0.03 |
| Jing | 0.15 | 0.94 | 0.14 | 0 | 0.01 | 0.04 | 0.03 | 0.02 | 0.05 | 0.04 | 0.03 | 0.05 | 0.02 | 0.02 | 0.04 | 0.03 |
| Donghai | 0.76 | 0.18 | 0.14 | 0.03 | 0.01 | 0.05 | 0.01 | 0 | 0.05 | 0.02 | 0.02 | 0.01 | 0.03 | 0.04 | 0.02 | 0.03 |
| Lingbi | 0.14 | 0.94 | 0.13 | 0.03 | 0.03 | 0.01 | 0.01 | 0.02 | 0.02 | 0.02 | 0.04 | 0.03 | 0.04 | 0.02 | 0.01 | 0.04 |
| Jingde | 0.16 | 0.77 | 0.13 | 0.04 | 0.03 | 0.03 | 0.03 | 0.01 | 0.01 | 0.04 | 0 | 0.04 | 0.01 | 0.01 | 0.01 | 0.01 |
| Huoqiu | 0.67 | 0.15 | 0.1 | 0 | 0.02 | 0.01 | 0.05 | 0.01 | 0.05 | 0.03 | 0.03 | 0.02 | 0.02 | 0.04 | 0.04 | 0.03 |

**Note: Interpretation of the Headers of Table 10**

| Symbol | Meaning |
|---|---|
| Ocr | Incidence of public service shortfalls |
| deg | Extent of shortfalls in basic public services |
| Ind | Shortboard index of basic public services |
| A | Digital economy index |
| B | Participation in the cooperative economy |
| C | Share of agricultural mechanization |
| D | Number of museums per 10,000 people |
| E | Number of cultural centers per 10,000 people |
| F | Urban greening coverage |
| G | Garbage disposal rate |
| H | Sewage treatment rate |
| I | Hospitals per 10,000 people |
| J | Beds per 10,000 people |
| K | Penetration rate of water supply |
| L | Public transportation & road density |
| M | Number of books per 10,000 people |

The weak county of Guannan and the relatively strong county of Donghai are both from Lianyungang City, Jiangsu Province. Similarly, the weak county of Jinzhai and the relatively strong county of Huoqiu are both from Lu'an City, Anhui Province. Both the findings confirm our previous conclusion that disparities within groups are the more pressing issue in the equalization of basic public services. For the weaker counties, special attention should be paid to the improvement and enhancement of their most disadvantaged short boards. Table 10 has comprehensively presented the most disadvantaged short boards for the 40 counties above in detail.

In addition, Table 10 also reveals that the shortcomings between different weak counties reflect both individualized differences and certain commonalities. For example, bridging the gap in the number of cultural centers per 10,000 people can effectively help Dingyuan, Jixi, Jinzhai, Susong and many other counties and municipalities to raise the level of basic public service equalization, while raising the level of participation in the cooperative economy, the digital economy index, and the penetration rate of water supply are individualized initiatives for Dingyuan, Jixi and Gunnan County respectively. As a result, when the higher-level government drafts targeted policies for the weak counties under its jurisdiction to improve the equalization level of basic public services, it should not only pay attention to the common problems of different weak counties, but also optimize them from a personalized perspective in accordance with the local conditions.

## 5.2 $\beta$ convergence model to measure the convergence in the development of basic public services

After using the AF model to accurately identify the remaining shortcomings in promoting the equalization of basic public services in the Yangtze River Delta region, we then continue to verify the convergence trend of the development of basic public services by using the $\beta$ convergence model (Kovács et al., 2022).

### 5.2.1 Introduction to $\beta$ convergence model

This paper uses the conditional $\beta$ convergence model to study the influencing factors affecting the growth of public service equalization and to explore whether there is a convergence trend in the level of public service equalization. Since the level of public service equalization varies between municipalities, this paper introduces control variables. Meanwhile, in order to eliminate the spatial correlation caused by the $\beta$ estimation bias, the convergence of equalization is tested for robustness and spatial correlation is incorporated, based on which the spatial lag model and spatial error model are established. The specific model construction process is as follows:

$$ln\frac{D_{i,t+1}}{D_{it}} = \beta lnD_{it} + \rho \sum_{i\neq j} w_{ij} ln\frac{D_{j,t+1}}{D_{jt}} + \sum_{k=1}^{L}\gamma_k X_{kit} + \mu_i + \lambda_t + u_{it}$$

$$ln\frac{D_{i,t+1}}{D_{it}} = \beta lnD_{it} + \sum_{k=1}^{L}\gamma_k X_{kit} + \mu_i + \lambda_t + u_{it}$$

$$u_{it} = \lambda \sum_{i\neq j} w_{ij}\gamma_{jt} + e_{it} \tag{11}$$

Here $\mu_i$ is the individual fixed effect, $\lambda_t$ is the time fixed effect, $u_{it}$ is the random perturbation term. $\beta$ is a key factor in the convergence analysis, and when $\beta$ is significantly negative, it indicates the existence of $\beta$ convergence, of which the speed of convergence $s = -\ln(1+\beta)T$, that is, the larger the absolute value of $\beta$, the faster the speed of convergence. $X_{jit}$ denotes the control variable $k$ of the city $i$ at time $t$, and $\gamma_k$ is the coefficient of the control variable $k$.

are utilized, i.e., total domestic economic activity, import and export trade, the level of development of the three major industries, the level of resource supply and the level of environmental pollution. The data are all the same as the coupled model. And the digital economy HP index is additionally introduced based on existing research.

Thereafter the regression and results were analyzed via Stata 17.0 software, using the modernization indicators in the coupling coordination model combined with the digital economy development index.

### 5.2.2 Analysis of β convergence model results

The $\beta$ conditional convergence test results are shown in Table 11. Considering the significant spatial positive autocorrelation of public service equalization and its dimensions, the $\beta$ convergence test uses the inverse distance spatial weight matrix reflecting the spatial distribution of cities with respect to distance. Both the LM-error and LM-lag tests for public service equalization and its dimensions rejected the original hypothesis of using no spatial error effect or no spatial autocorrelation effect at the 0.01 significance level, and when using the robust LM-error test and the robust LM-lag test, the robust LM-error test turned out more significant, so the spatial error model was used as a baseline for establishing the conditional convergence model.

Table 11. β Conditional Convergence Test Results

| Variables | Equalization of Public Services | |
|---|---|---|
| | SEM | OLS |
| R-squared | 0.265 | 0.725 |
| R-LM-Lag | 35,651*** | — |
| LM-lag | 13.901*** | — |
| LM-error | 85,653*** | — |
| R-LM-error | 135,654*** | — |
| λ | 2.541***(156.891) | — |
| β | -0.565 | -0.465 |
| s | 0.086 | 0.070 |
| Time fixed effect | yes | yes |

| Individual fixed effect | yes | yes |
|---|---|---|
| Digital economy development index | -0.112 | -0.132 |
| GDP | -0.04 | -0.037 |
| Import and export trade | -0.044 | -0.056 |
| Level of development of the primary sector | 0.033 | 0.038 |
| Level of development of the secondary industry | 0.042 | 0.046 |
| Level of development of the tertiary industry | 0.039 | 0.041 |
| Level of resource availability | 0.326 | 0.387 |
| Level of environmental pollution | 0.234 | 0.265 |

First we focus on the preliminary analytical conclusions as follows:

**a.** After eliminating spatial autocorrelation, the speeds of convergence in all dimensions of public service equalization all slightly increased.

**b.** Observing the level of basic public service equalization from the 41 municipalities studied, all the dimensions have a 0.01 significance level of $\beta$ convergence, i.e. regions with lower levels of public service equalization have faster growth rates and are expected to converge to the same level of basic public services.

**c.** After introducing control variables, the level of public service equalization and the speed of convergence of each dimension both increase, indicating that other variables have a positive effect on promoting the convergence of public service equalization.

**d.** The $\lambda$ of public service equalization and its dimensions is significantly positive, showing that public service equalization and its dimensions have significant spatial spillover effects, which can increase the speed of convergence of the level of basic public service equalization in the areas above.

Based on preliminary conclusions, this paper then delivers some in-depth analyses of conclusions presented.

**a.** The rise in the economic level of cities has not significantly increased the growth rate of the equalization level of basic public services, indicating that in the process of economic development, there is an imbalance of distribution within municipalities, which is prone to fall into the dilemma of polarization under the high level of development.

**b.** With the expansion of import and export trade, the growth rate of public service equalization has slowed down, indicating that the efficiency of fiscal expenditure in regulating regional synergistic development and income distribution is still relatively low, thus not effectively bridging the development differences among counties or between urban and rural areas.

**c.** The increase in the level of resource supply and the decrease in the level of environmental pollution both show a significant positive effect on the increase in the level of equalization of basic public services. It shows that the government can improve the supply of public services from the aspects of resources and environment.

**d.** The growth of the digital economy significantly inhibits the growth of public service equalization. Even though digital economy is widely believed to have certain benefits for the overall welfare of societies, this paper, through empirical analysis, argues that the growth of digital economy as a skill-intensive industry does have an exclusionary effect on low-skill level labor, meaning that while digital economy raises the technological level in the region, it will in turn inhibit a portion of low-skill level labor force from being employed. This dilemma needs further tradeoff.

## 6. Conclusions from the targeted study of the Yangtze River Delta region

Since 2017, the dual systems of basic public services and the degree of modernization have been in a coupled and coordinated development pattern, and have kept a steady upward trend except for a significant decline in 2020 (probably due to the shock of Covid-19 pandemic). It can be assumed that advancing the level of basic public services contributes to the process of regional modernization.

Meanwhile, according to the calculation and analysis of Moran Index, it can be found that the overall basic public service level of each region shows positive correlation during the five years, while the higher or lower the level of basic public service equalization, the more likely the aggregation phenomenon is to happen. But at the same time, the degree of aggregation shows a decreasing trend over time, and locally the low-low aggregation area gradually decreases, followed by the phenomenon of high-high and high-low aggregation areas disappearing in 2021, indicating that the level of basic public service equalization gradually increases.

In addition, the measurement of the Theil index shows that the regional differences in the level of basic public services in the Yangtze River Delta region have a narrowing trend in general, especially the larger decline in 2019. Specifically, in the four provinces or cities of Jiangsu, Zhejiang, Shanghai and Anhui, the differences within groups are more significant than those between groups, which reminds provincial governments that they should pay more attention to the problem of the imbalance in the development of basic public services in different regions within the province.

And through the AF method to analyze the 40 counties under the jurisdiction of the 10 cities with the lowest five-year average development level of basic public services, it is found that there are a total of 13 counties with weak public service problems, and on average, there are 4.8 indices of shortboard problems in each county with weak public service. Observing the specific values of shortboard index of each region, we advocate paying more attention to the short boards with high shortcoming indices, such as improving the participation in cooperative economy and the number of hospitals per 10,000 people, etc., for Dingyuan County, in order to improve the overall level of basic public services in a more effective manner.

Finally, through the calculation of the $\beta$ convergence model, regions with lower levels of public service equalization have faster growth rates and have the potential of converging to the same level of basic public services in the long run.

## 7. Recommendations for developing countries to enhance public service equalization and modernization from the lesson of YRD region

Achieving the equalization of public services is a huge step forward in the modernization of developing countries. From the analysis of China's Yangtze River Delta region, this paper attempts to distill some universal initiatives to promote the regional equalization of basic public services, providing policy suggestions for other regions or countries to promote the balanced development of public services & modernization, and finally to realize high-quality development.

## 7.1 Focusing on shortboards and strengthening regional linkages

On the one hand, the impact of the level of basic public services on people's living standards has a certain "barrel effect", indicating that to strengthen the supply of basic public services, policymakers should not merely focus on the development of a more balanced "long board". Actually, more efforts need to be put into the provision and advancement of those relatively week aspects of basic public services at the current stage, in order to meet people's needs more effectively. In other words, starting from the "short boards" is more likely to bring in targeted optimization, thus enhancing the overall level of basic public services by maximizing the marginal rate of return. On the other hand, basic public services often involve both intra- and inter-regional collaboration. So when promoting the construction of basic public services within its jurisdiction, the higher-level government should implement more targeted policies towards those relatively weak cities and counties, not only solving common problems, but also focusing on local conditions and providing what weak counties really need, so as to promote the regional equalization of basic public services.

## 7.2 Emphasizing the quality of development and preventing existing gaps from expansion

Through the $\beta$ conditional convergence test, we find that the increase in the level of urban economic development has not been able to significantly increase the growth rate of the equalization level of basic public services. At the same time, with the expansion of import and export trade, the growth rate of public service equalization has even slowed down. Therefore, we can not simply rely on economic growth to promote the equalization level of basic public services, but should effectively leverage the function of financial transfers from the perspective of regional synergistic development, whilst focusing on the control of existing urban-rural, inter-county and intra-county differences in public services. In particular, during the post-epidemic era, it is exceedingly important to avoid possible reversal in the level of equalization of basic public services as the economy rebounds, aiming at steadily promoting high-quality development.

## 7.3 Guiding emerging industries and stimulating regional vitality

Based on the previous analysis, we find that the advancement of emerging industries, especially skill-intensive ones like digital economy, is likely to hinder the growth of the equalization level of basic public services. For this reason, when introducing high-tech industrial parks and other kinds of regional complexes, the government should pay attention to their spillover effects on the surrounding areas, trying to maximize their positive impacts. To be specific, on the one hand, government could stimulate the transformation and upgrading of the surrounding areas' industries through the introduction of new technologies and business models to. And on the other hand, the surrounding areas may improve their corresponding public service provision through the introduction and development of the non-core functions of newly built industrial parks, so as to shape a two-way synergy to maximize the positive spillover effect, and strive to make the new technologies become a vibrant kind of vitality that empowers public service advancement, instead of inhibiting and hindering it.

## Statements & Declarations

### Funding

The authors declare that no funds, grants, or other support were received during the preparation of this manuscript.

### Competing Interests

The authors have no relevant financial or non-financial interests to disclose.

### Author Contributions

The authors contributed equally to the study conception and design. After the authors' collaboration on material preparation, data collection and analysis, both the authors read and approved the final manuscript.

### Acknowledgement

We are grateful to the previous field investigation programs conducted by our colleagues. Their experiences and findings provided us with a source of inspiration before drafting this manuscript. The authors are responsible for all remaining errors.